\documentclass[aps,twocolumn,prd,showpacs,nofootinbib,showkeys,superscriptaddress,preprintnumbers]{revtex4} 

\usepackage{hyperref}
\usepackage{latexsym}
\usepackage{epsfig}
\usepackage{graphicx,subfigure}
\usepackage{ulem}
\usepackage{color}
\usepackage[utf8]{inputenc}

\begin{document}

\newcommand{\tcb}{\textcolor{blue}}

\def\ga{\mathrel{\raise.3ex\hbox{$>$\kern-.75em\lower1ex\hbox{$\sim$}}}}
\def\la{\mathrel{\raise.3ex\hbox{$<$\kern-.75em\lower1ex\hbox{$\sim$}}}}

\def\be{\begin{equation}}
\def\ee{\end{equation}}
\def\bea{\begin{eqnarray}}
\def\eea{\end{eqnarray}}

\def\betap{\tilde\beta}
\def\del{\delta_{\rm PBH}^{\rm local}}
\def\Msun{M_\odot}
\def\Rcl{R_{\rm clust.}}
\def\fPBH{f_{\rm PBH}}

\def\JGB#1{\textcolor{red}{\tt[JGB: #1]}}

\newcommand{\dd}{\mathrm{d}} 
\newcommand{\Mpl}{M_P} 
\newcommand{\mpl}{m_\mathrm{pl}} 
 
\thispagestyle{empty}
 
\title{GW190425, GW190521 and GW190814:  \\ 
Three candidate mergers of primordial black holes from the QCD epoch} 

\author{Sebastien Clesse}\email[]{sebastien.clesse@ulb.ac.be}
\affiliation{Service de Physique Th\'eorique, Univesrit\'e Libre de Bruxelles (ULB), Boulevard du Triomphe, CP225, B-1050 Brussels, Belgium}

\author{Juan Garc\'ia-Bellido}
\email[]{juan.garciabellido@uam.es}
\affiliation{Instituto de F\'isica Te\'orica UAM/CSIC, Universidad Aut\'onoma de Madrid, 28049 Madrid, Spain}

\date{\today}
\begin{abstract}  
The two recent gravitational-wave events GW190425 and GW190814 from the third observing run of LIGO/Virgo have both a  companion which is unexpected if originated from a neutron star or a stellar black hole, with masses $[1.6-2.5]~\Msun$ and $[2.5-2.7]~\Msun$ and merging rates $ 460^{+1050}_{-360} $ and $ 7^{+16}_{-6}$ events/yr/Gpc$^3$ respectively, at 90\% c.l.. Moreover, the recent event GW190521 has black hole components with masses 67 and $91~\Msun$, and therefore lies in the so-called pair-instability mass gap, where there should not be direct formation of stellar black holes. The possibility that all of these compact objects are Primordial Black Holes (PBHs) is investigated.   The known thermal history of the Universe predicts that PBH formation is boosted at the time of the QCD transition, inducing a peak in their distribution at this particular mass scale, and a bump around $30-50~\Msun$.  We find that the merging rates inferred from GW190425, GW190521 and GW190814 are consistent with PBH binaries formed by capture in dense halos in the matter era or in the early universe.  At the same time, the rate of black hole mergers around $30~\Msun$ and of sub-solar PBH mergers do not exceed the LIGO/Virgo limits. Such PBHs could explain a significant fraction, or even the totality of the Dark Matter, but they must be sufficiently strongly clustered in order to be consistent with current astrophysical limits.
\end{abstract}

\pacs{04.70.Bw, 97.60.Lf, 95.35.+d}

\maketitle

\textit{Introduction:}  The detection in 2015 by Advanced LIGO of the gravitational waves (GWs) emitted during the final phase of the merging of two black holes (BHs)~\cite{Abbott:2016blz}, has been an incredible tour de force rewarded by the 2017 Nobel Prize in Physics.  GW observations open a new window to study BH formation scenarios and to test fundamental physics.
The first series of GW observations by LIGO/Virgo have brought their share of surprises, like progenitor masses above expectations, suggesting that they may come from low-metallicity environments if they are of stellar origin, and low effective spins that are hard to explain in standard stellar evolution scenarios~\cite{Belczynski:2015tba}.  Recently, the two events GW190425 and GW190814, which had no electromagnetic counterpart, have revealed the existence of compact objects of masses between $1.8$ and $2.7~\Msun$~\cite{Abbott:2020uma,Abbott:2020khf}.  
This is above the mass of all known binary neutron stars~\cite{Ozel:2016oaf}, and below expectations for stellar black holes, in the so-called lower mass gap, see however~\cite{Gupta:2019nwj}.  The existence of black holes in this range of masses is further supported by  a recent microlensing survey towards the galactic bulge, based on observations by OGLE and Gaia~\cite{Wyrzykowski:2019jyg}.  Furthermore, GW190814 is an unequal mass binary merger, with a mass ratio of about $q=0.1$.  The spin of its primary component is the best constrained so far and it is very low, 
$|\chi_1| < 0.07$~\cite{Abbott:2020khf}.
The inferred merging rate, $\tau = 7^{+16}_{-6}$ events/yr/Gpc$^3$ is only slightly below that for massive black holes holes and seems to be  incompatible with current astrophysical models~\cite{Abbott:2020khf,Zevin:2020gma}. 
Moreover, the recent event GW190521 has black hole components with masses $67$ and $91~\Msun$~\cite{Abbott:2020tfl}, and therefore lies in the upper pair-instability mass gap where eventual black holes should not have directly formed by stellar explosions. All this suggests the need of revising and improving stellar or black hole evolution scenarios, or of seriously considering the existence of a new population of black holes of primordial origin~\cite{Clesse:2017bsw}.

\textit{Primordial Black Holes}  (PBHs) may have formed in the early Universe due to the gravitational collapse of pre-existing, order one density fluctuations~\cite{Hawking:1971ei,Carr:1974nx,1975Natur.253..251C}. These can take their origins in the inflationary era~\cite{Carr:1994ar,Ivanov:1994pa,Randall:1995dj,Dolgov:1992pu,GarciaBellido:1996qt} and in some models, a wide distribution of stellar-mass PBHs can be produced~\cite{2015PhRvD..92b3524C,Garcia-Bellido:2017mdw,Ezquiaga:2017fvi,Chapline:2018sbr} at the epoch of the QCD transition.  These may contribute to a fraction, or even the totality of the Dark Matter (DM) in the Universe.  
PBHs would naturally have low spins~\cite{DeLuca:2019buf}, different from the predictions of stellar models~\cite{Belczynski:2015tba,Fuller:2019sxi}, see however~\cite{DeLuca:2020bjf}.

Soon after the first gravitational-wave detection, it has been suggested that the progenitors of GW150914 were PBHs~\cite{Bird:2016dcv,Clesse:2016vqa,Sasaki:2016jop}, see also~\cite{Kashlinsky:2016sdv} for the possible connection with cosmic infrared background anisotropies.   Two binary formation channels have been investigated:  by capture in dense halos~\cite{Bird:2016dcv}, such as ultra-faint-dwarf-galaxies~\cite{Clesse:2016vqa}, or right after their formation as a result of the Poissonian fluctuations in their initial positions~\cite{Sasaki:2016jop}.   Both channels can lead to merger rates compatible with LIGO/Virgo observations if PBHs contribute with a significant fraction of the DM~\cite{Clesse:2017bsw,Carr:2019kxo,DeLuca:2020qqa,Jedamzik:2020ypm}.\footnote{For primordial binaries, a higher rate was obtained by Sasaki et al.~\cite{Sasaki:2016jop} but since then N-body simulations have shown a rate suppression if $f_{\rm PBH} \gtrsim 0.1 $~\cite{Vaskonen:2019jpv} due to tidal disruption by early-forming PBH clusters, possibly down to a value compatible with gravitational-wave observations~\cite{Raidal:2018bbj}.}  

\begin{figure}[t]
    \centering
    \includegraphics[width=8.7cm]{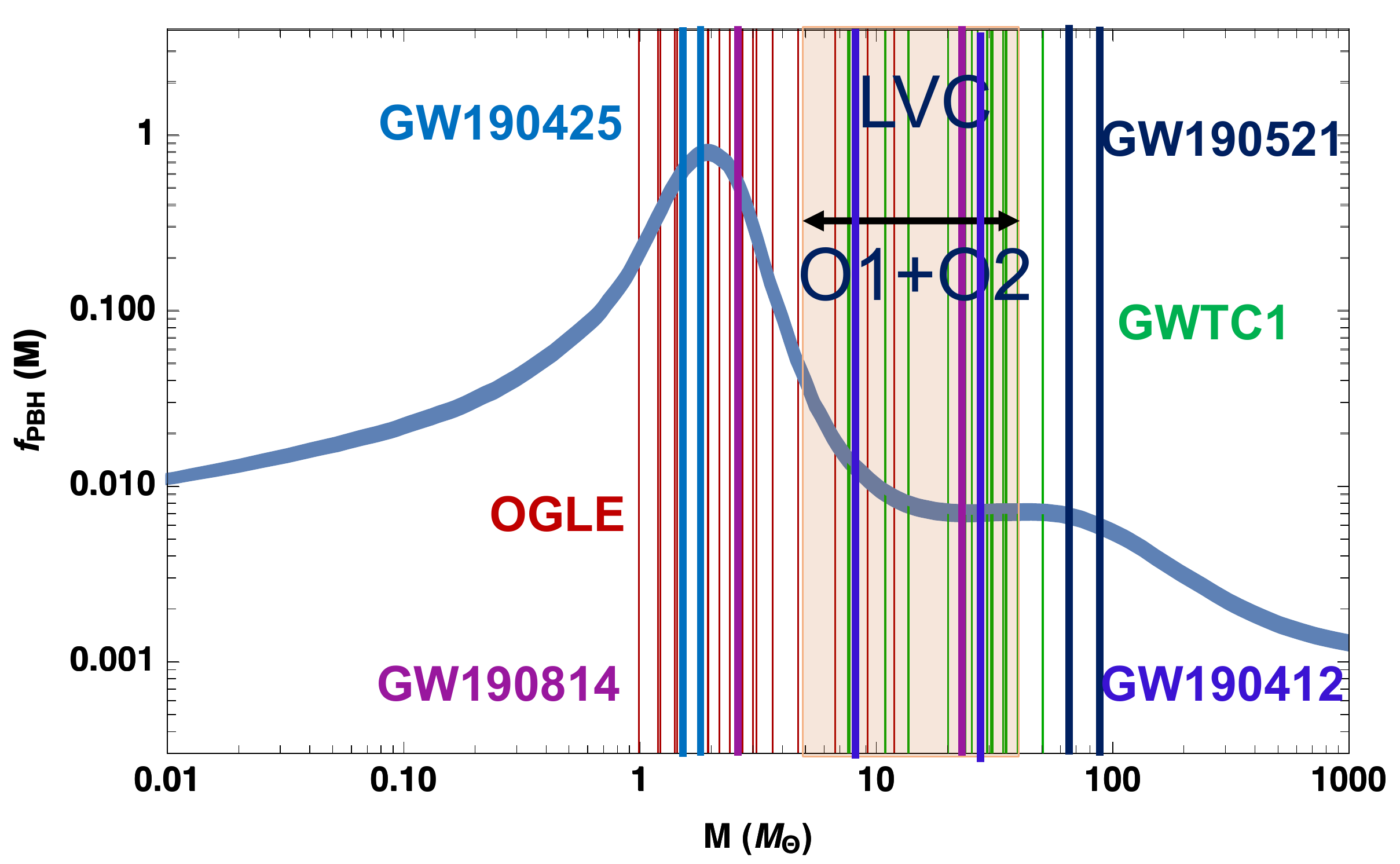}
    \caption{The PBH mass distribution (blue line) from an almost scale-invariant primordial power spectrum of curvature fluctuations with $n_{\rm s} = 0.97$, normalized such that $f_{\rm PBH} = 1$, with a PBH/Hubble mass ratio of $\gamma = 0.8$ and showing the features induced by the QCD transition in the form of a proton-peak at $2-3~\Msun$ and a pion-bump at $30-50~\Msun$. Colored vertical lines indicate the best-fits for the component masses of GW190412, GW190425, GW190521 and GW190814. We also show the GWTC1 events (green vertical lines), and the compact lens masses (red lines) measured in microlensing events by OGLE and GAIA towards the galactic bulge.}
    \label{fig:fPBH}
\end{figure}

The limits on the PBH abundance from various astrophysical and cosmological probes (for a review or perspectives, see e.g. Refs~\cite{Carr:2009jm,Carr:2016drx,Carr:2020gox,Carr:2020xqk,Ali-Haimoud:2019khd} and references therein) exclude monochromatic PBH models, but these are anyway unrealistic from the theoretical point of view, because any statistical distribution of pre-existing inhomogeneities leads to the formation of PBHs near the regime of critical collapse, and thus to a peaked mass function of non-negligible width~\cite{Niemeyer:1997mt,Musco:2008hv,Musco:2012au}.  At the solar mass scale, only lensing constraints seem to exclude PBHs to constitute a large fraction of the DM, see however~\cite{Garcia-Bellido:2017xvr,Calcino:2018mwh}.  But wide mass functions should also change the clustering properties of PBHs~\cite{MoradinezhadDizgah:2019wjf}, and if most of them are regrouped in dense halos whose size is only limited from below due to the process of dynamical heating, then those PBH clusters also act as a lens~\cite{Carr:2019kxo}.  This way, distant point sources become Einstein arcs and the magnitude of eventual microlensing events is damped below the detectable level. Other limits on the PBH fraction apply to lower or high mass scales, and so it is still plausible, even if debated and controversial, that an extended mass function and clustered PBHs around $2~\Msun$ constitute most of the DM in the Universe~\cite{Clesse:2017bsw,Carr:2019kxo}.  Such PBHs would have formed exactly at the time of the QCD transition, when quarks and gluons condensed into protons and neutrons.

\

\textit{The QCD transition} induces a temporary reduction of the equation of state of the Universe.  As a consequence of the exponential dependence of the PBH abundance on the equation of state (through the overdensity threshold value leading to gravitational collapse), PBH formation must have been boosted at the QCD transition~\cite{Jedamzik:1996mr,2018JCAP...08..041B,Carr:2019kxo}. Since this relies on known physics, it inevitably introduces a peak in the PBH mass function around the solar mass scale, as well as a bump around $30~\Msun$ corresponding to the moment when pions annihilate~\cite{2018JCAP...08..041B,Bianchi:2018ula,Carr:2019kxo}.  At the same time, the collapsing inhomogeneities into PBHs provide all the ingredients for an efficient baryogenesis, without the need to go beyond the standard model of particle physics~\cite{Garcia-Bellido:2019vlf,Carr:2019hud}.  In this scenario, if PBH are all of the DM or an important fraction of it, their abundance at formation is naturally connected to the baryon to photon ratio of the Universe.  The relative abundance of baryons compared to DM in the form of PBHs also suggests a ratio $\gamma \simeq 0.8$ between the PBH mass and the mass of the collapsing horizon-sized region. Different  assumptions related to the PBH collapse lead to a plausible range of $\gamma$ between 0.2 and 1~\cite{Byrnes:2018clq}.  In particular, a value around 0.2 would be motivated by a calculation based on the turn-around scale.  Ultimately, a more precise value of $\gamma$ should be computed from simulations of PBH formation in numerical relativity while taking into account the variation of the equation of state during the whole process of PBH fomation.  We consider here $\gamma = 0.8$ as a benchmark, but our conclusions remain valid for values between $0.6$ and $1$.  For lower values, the QCD-proton peak is shifted towards lower masses and becomes inconsistent with the rate limits imposed by LIGO/Virgo in the neutron star mass range.  Besides being plausible, $\gamma \approx 0.8$ is observationallly motivated by OGLE observations of a series of microlensing events towards the galactic center~\cite{Wyrzykowski:2019jyg} that have revealed the possible existence of an unexpected black hole population in the mass gap between $2$ and $5\, M_\odot$, which could be PBHs~\cite{Garcia-Bellido:2019tvz}. 

\textit{The PBH mass distribution} imprinted by the QCD transition for a nearly scale invariant power spectrum of curvature fluctuations with a spectral index of $n_{\rm s} = 0.97$ and normalized to get an integrated PBH abundance equal to the one of DM, is represented on Figure~\ref{fig:fPBH}.  The values of $n_{\rm s}$ compatible with the astrophysical limits on the abundance of PBHs are quite restricted, between $0.96$ and $0.98$ such that the mass function in the stellar mass range is relatively well defined~\cite{Carr:2019kxo}.  Almost scale invariance is a generic prediction of inflation.  Effectively, the model also describes well other scenarios with a broad power spectrum peak~\cite{2015PhRvD..92b3524C,Garcia-Bellido:2017mdw,Ezquiaga:2017fvi}.  This scenario therefore provides a strong motivation to search for PBHs between $2$ and $3~\Msun $ from the QCD-proton peak, eventually merging with PBHs from the QCD-pion bump.

\

\begin{figure}[t]
    \centering
    \includegraphics[width=9cm]{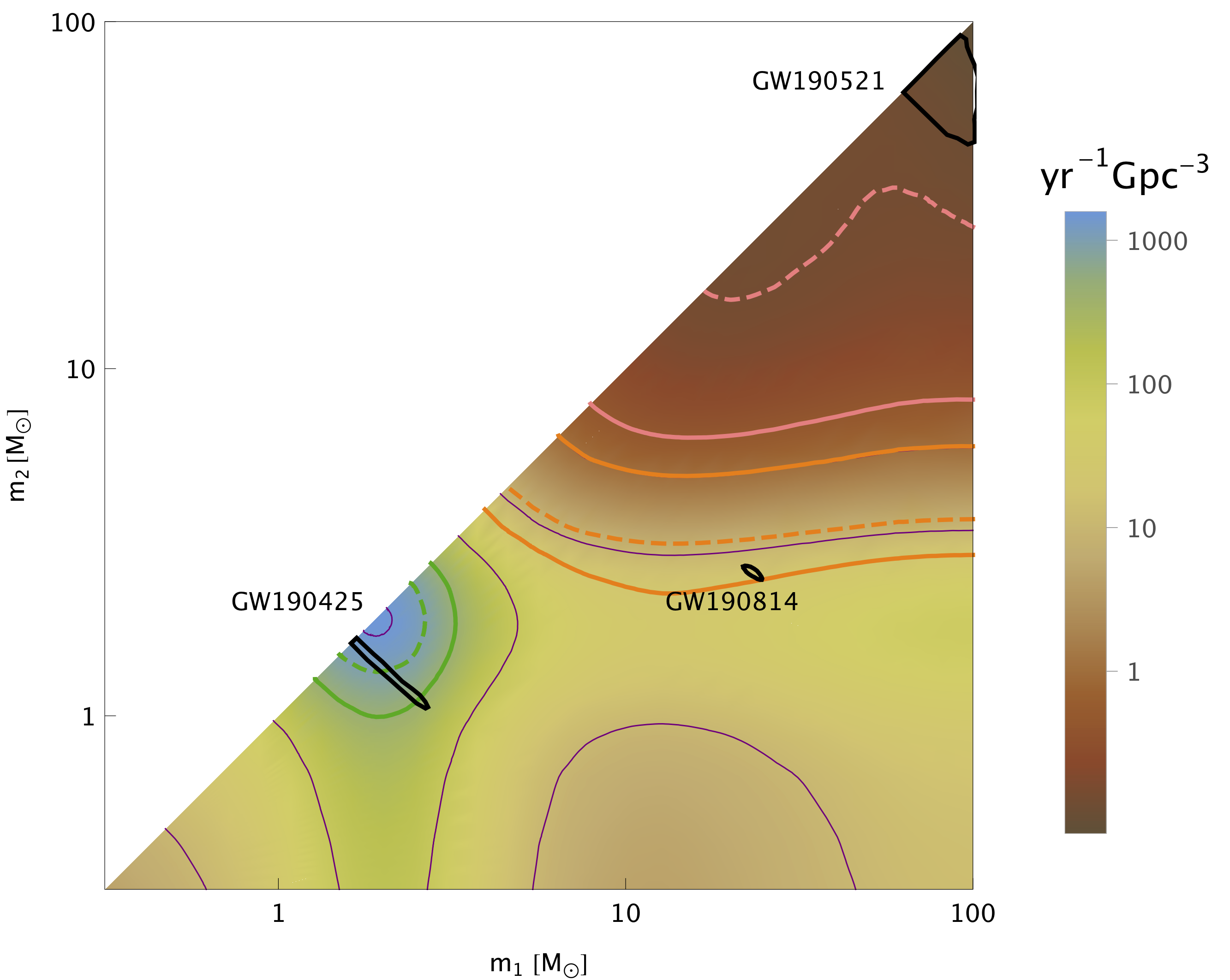}
    \caption{Merging rate distribution for the PBH mass distribution shown in Figure \ref{fig:fPBH}, for PBH binaries formed by tidal capture in dense clusters, with $R_{\rm clust} = 420 $.  The orange, green and pink isocontours correspond to the rates inferred from GW190814, GW190425 and GW190521 respectively, the solid lines corresponding to the $90\%$ confidence intervals and the dashed lines to the best fit.  The black lines represent the contours  ($90\%$ confidence limits) for the component masses of these events and all coincide with the rate predictions in our PBH scenario. }
    \label{fig:rates}
\end{figure}

\textit{GW190425  and  GW190814} both involve at least one compact object of this mass and may be a strong hint of their primordial origin.  GW190425 could thus be due to two PBHs from the QCD-proton peak, while GW190814 could involve one PBH from this peak and another one from the QCD-pion bump.  In order to test this hypothesis, the merging rates inferred by the LIGO/Virgo collaboration for these two events is a good discriminator.  For GW190425, it is evaluated at
$M\approx2.5~M_\odot$~\cite{Abbott:2020uma}, while for GW190814, at $M\approx2.7~M_\odot$~\cite{Abbott:2020khf}.  These rates should also be compared to the merging rate of heavier black hole mergers, since their suspected low spins may also be explained by a primordial origin~\cite{Garcia-Bellido:2020pwq}.  Note that it is extremely difficult  to explain the existence of binaries of astrophysical black holes in the mass gap, with low mass ratios and negligible spins.  It is unlikely that one component is a BH of stellar origin or a neutron star and that the other component is a PBH~\cite{Vattis:2020iuz,Tsai:2020hpi}. Moreover, the absence of tidal deformations in the waveform of GW190814, as well as the amount of total GW emitted versus the final mass of the merged black hole, may be a hint of a binary black hole (BBH), but one cannot exclude a neutron star - black hole (NSBH) binary~\cite{Abbott:2020khf}.

\

\textit{GW190521} involves two massive black holes that fall in the middle of the upper mass gap ($[60 - 120]\,\Msun$)~\cite{Abbott:2020tfl,Abbott:2020mjq} coming from (pulsating) pair-instability supernovae, where such black holes should not form.  As a possible explanation, one can invoke hierarchical mergers, a scenario in which each of these black holes originates from the previous merging of two $[30-40] M_\odot$ black holes.  However, for this explanation to be viable, one needs dense environments where black hole mergers are frequent, as well as kick velocities from the merger that do not exceed the escape velocity of this environment.  Finally, it is statistically unlikely to detect first  the merging of two such black holes rather than the merging of one of them with a $30 M_\odot$ black hole~\cite{Fishbach:2020qag}.  Alternatively, their mass and merging rate may suggest that their origin is primordial. In fact, their mass lie precisely where there is a bump due to pion annihilation in the thermal history scenario of Ref.~\cite{Carr:2019kxo}.  Hereafter we show that the merging rates of such PBHs are consistent with the rate inferred from the observation of GW190521.

\

\textit{PBH merging rates:}  PBH binaries can form by tidal capture in dense halos in the matter era, or in the early universe before the matter-radiation equality, if they formed sufficiently close to each other for their dynamics to decouple from the expansion of the Universe.  Each channel gives a specific mass-dependence of the merging rate.  

For PBHs in dense halos, the merging rate distribution is given by~\cite{Clesse:2016vqa}
\bea 
    \frac{\dd\tau}{\dd \ln m_1 \,\dd \ln m_2 } & = & R_{\rm clust.} \times f(m_1) f(m_2)  \nonumber \\
    & \times & \frac{(m_1 + m_2)^{10/7}}{(m_1 m_2)^{5/7}} \rm{yr^{-1}Gpc^{-3}}, 
     \label{eq:ratescatpure2}
\eea
where $\Rcl$ is a scaling factor that depends on the PBH clustering properties, including their velocity distribution, $f(m_{\rm PBH}) \equiv \dd \rho_{\rm PBH} / \dd \ln m_{\rm PBH} $ is the PBH mass distribution represented on Figure~\ref{fig:fPBH}, $\rho_{\rm PBH}$ is the cosmological density of PBHs today, and $m_1$, $m_2$ are the two merging black hole masses.  Halo mass functions compatible with the standard $\Lambda$CDM cosmological scenario typically lead to $R_{\rm clust.}\approx 1-10 $~\cite{Bird:2016dcv}.  For our mass distribution, this is too low to explain the merging rate inferred from GW190425, at which mass one has $f(m_{\rm PBH}) \simeq 1$.  This is also too low to explain the rate at larger mass, inferred from other black hole mergers~\cite{Abbott:2016nhf,LIGOScientific:2018jsj}.  However, a wide mass distribution naturally leads to enhanced clustering~\cite{MoradinezhadDizgah:2019wjf} for several reasons.
On the one hand, because of the existence of initially large curvature fluctuations on scales smaller than the ones probed by the cosmic microwave background and large scale structures.  On the other hand, because intermediate and supermassive PBHs act as an additional seed of structures~\cite{Rubin:2001yw,Khlopov:2002yi,Clesse:2015wea,Carr:2018rid}.  But the most important source of clustering comes the additional Poisson term in the power spectrum, as discussed in Appendix.  Finally, in some scenarios PBH may have directly formed in clusters~\cite{Khlopov:2004sc,Inman:2019wvr}.  The importance of these effects is strongly model dependent and still an open question.  But due to the discrete nature of PBHs, dense clusters are subject to dynamical heating~\cite{Trashorras:2020mwn}.  Typically PBH clusters of radius less than a parsec are dynamically unstable and expand, up to the scale of ultra-faint-dwarf-galaxies~\cite{Clesse:2016vqa}. In Appendix, we show that these effects provide strong theoretical motivations for $R_{\rm clust} \sim 10^2$.   Clustering is also required in order to evade the microlensing limits on the PBHs~\cite{Garcia-Bellido:2017xvr,Calcino:2018mwh,Carr:2019kxo,Belotsky:2018wph}.  We find a range $R_{\rm clust} = [400 - 450] $, in such a way that the integrated merging rate for a primary mass $m_1 >5 M_\odot$ and mass ratios  $q \equiv m_2/ m_1 > 0.2$, is around $20 \, {\rm yr}^{-1}{\rm Gpc}^{-3}$.  This is compatible with the limits from LIGO/Virgo observations~\cite{LIGOScientific:2018jsj,Abbott:2016nhf} but also means that PBH binaries would likely constitute a sub-dominant fraction of the merger rate observed above $20~\Msun$, with a larger fraction due to stellar BH binaries. The resulting merger rate distribution is shown on Figure \ref{fig:rates}, together with isocontours corresponding to the rate values (best fit and 90\% c.l.) inferred from GW190425, GW190521 and GW190814~\cite{Abbott:2020uma,Abbott:2020mjq,Abbott:2020khf}.  These are perfectly consistent within the 90\% c.l.  for the two compact object masses.   Additionally, one can notice that GW190425, GW190521 and GW190814 lie in the three regions with a higher expected detection rate when one takes into account the detector sensitivity identified in Ref.~\cite{Carr:2019kxo}.  Above $15 \, M_\odot$, the rate distribution is effectively well approximated by Model B of~\cite{LIGOScientific:2018jsj}, assuming $\dd \tau / \dd m_1 \propto m_1^{-\alpha} q^{\beta_q}$. We find that $\alpha \approx 1$ that is consistent with the observations of the second observing run of LIGO/Virgo.  Nevertheless, we also find that $\beta_q \approx -1$, a value disfavored by observations.  One should however notice that Model B with $\beta_q >0$ is ruled out by GW190814 and so a more detailed Bayesian analysis would be in favor of the PBH model.
Finally, Figure~\ref{fig:rates} shows the merger rate distribution for both sub-solar PBHs and more massive ones.  It is consistent with the limits imposed by the search of such objects in the second observing run of LIGO/Virgo~\cite{Authors:2019qbw,Phukon:2021cus}.  It also motivates an extension of this search to sub-solar PBHs with a companion of mass larger than $2~\Msun$, which have a total rate of $\tau \approx 200 \, {\rm yr}^{-1}{\rm Gpc}^{-3}$.

Therefore, a PBH scenario taking into account the thermal history with binaries formed by tidal capture in halos, could explain at the same time the mass, spins and rate of the three unexpected events GW190425, GW190521 and GW190814, while being consistent with rate limits at large masses and at sub-solar masses. 

\begin{figure}
\vspace{5mm}
    \centering
    \includegraphics[width=9cm]{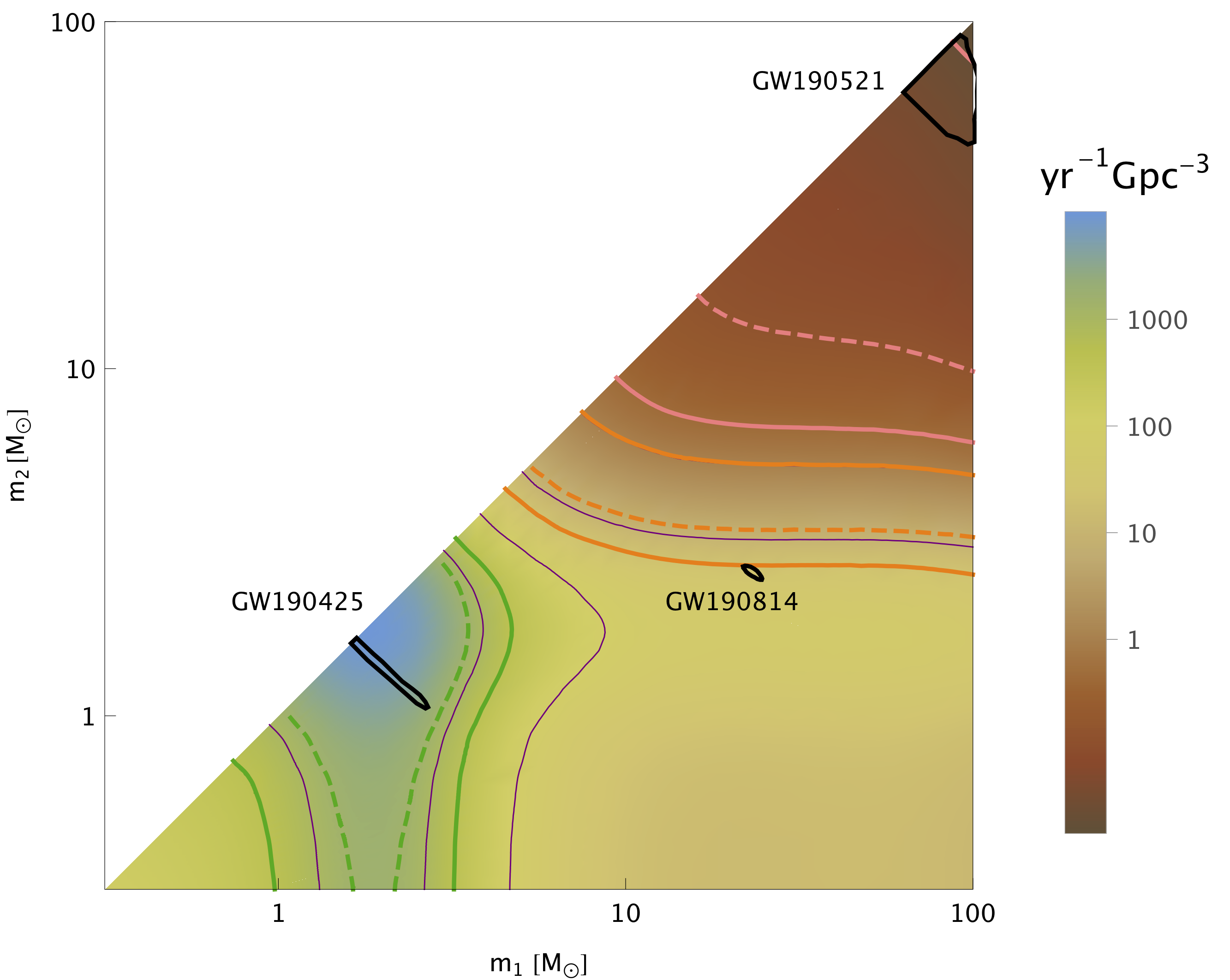}
    \caption{Same as Figure~\ref{fig:rates}, but for PBH binaries formed by tidal capture in the radiation era, assuming a suppression factor $f_{\rm sup} = 0.0025$ or, alternatively, $f_{\rm PBH} = 0.05$.  }
    \label{fig:ratesprim}
\end{figure}

One can also examine if PBH binaries formed by tidal capture in the early Universe~\cite{Sasaki:2016jop} can explain those merging rates.  These can be found when PBHs are generated sufficiently close to each other, as a result of their Poissonian spatial separation at formation.  The gravitational influence of one or several PBHs nearby  prevent the two black holes to merge directly and instead form a binary.  Eventually, the binary is sufficiently stable and it takes of the order of the age of the Universe for the two black holes to merge.  If one assumes that early forming PBH clusters do not impact the lifetime of those primordial binaries (a criterion satisfied for $f_{\rm PBH} \lesssim 0.1 $~\cite{Vaskonen:2019jpv}), the present merging rate is approximately given by~\cite{Raidal:2018bbj,Gow:2019pok,Kocsis:2017yty} 
\bea 
         \frac{\dd\tau}{\dd \ln m_1 \,\dd \ln m_2} &\approx & 1.6 \times 10^6 \,{\rm Gpc^{-3} yr^{-1}} f(m_1) f(m_2) f_{\rm sup}   \nonumber \\  
         & \times & \left( \frac{m_1 + m_2}{M_\odot}\right)^{-\frac{32}{37}} \left[\frac{m_1 m_2}{(m_1+m_2)^2}\right]^{-\frac{34}{37}} ~ , \label{eq:tauEB}
\eea
If PBHs contribute predominantly to the DM, we effectively describe the above mentioned effect by including in the previous equation a suppression factor $f_{\rm sup}$ whose possible value is still rather unclear and can depend on numerous effects.  We argue in appendix A that $f_{\rm sup} = 0.0025$, independent of the PBH mass, which reproduces a rate of $\tau \approx 20 $ events/yr/Gpc$^3$, as for PBH binaries formed in clusters.  We show in Appendix that such a value for $f_{\rm sup}$ is motivated by N-body simulations and the most recent analytical prescriptions for the rate suppression.  Alternatively, one can consider a rescaled mass function giving $f_{\rm PBH} = 0.05$ with no suppression.  We find that the rates for GW190814 and GW190425 can be explained by early binaries as well, as shown in Figure~\ref{fig:ratesprim}.  Nevertheless, for this formation channel, the rates at larger masses is reduced by up to one order of magnitude, and therefore explaining at the same time the GW events observed in the second observing run appears to be challenging.  For the same reason, for the component masses of GW190521, the model predicts a merging rate that is still compatible but near the $90\%$ lower limit of the inferred rate.   Finally, we found that the merging rate of sub-solar binaries is of order $\tau \approx 900$ events/yr/Gpc$^3$ if $m_1 < 2 M_\odot$, consistent with current limits, and $\tau \approx 430$ events/yr/Gpc$^3$ if $m_1 > 2 M_\odot$.

\

\textit{Inferring the PBH abundance:} In order to compare
the rate in Eq.~(\ref{eq:ratescatpure2}) with the actual observations we need to fix the scaling factor $\Rcl$, which depends on both the PBH fraction $\fPBH$ and their clustering properties as a function of redshift. There are recent analysis of merger rates due to clustering PBHs~\cite{Raidal:2018bbj,Jedamzik:2020ypm} which show that three-body encounters inside dense clusters rather than increasing the rate of events actually reduce them due to the breaking of those binaries~\cite{Trashorras:2020mwn}. At the end, the rate is compatible with that observed by LIGO if all of the DM is composed of PBH. The usual constraints on monochromatic mass distributions of PBH uniformly distributed in space no longer apply~\cite{Calcino:2018mwh,Garcia-Bellido:2018leu,Carr:2020xqk}, and the clustered wide mass distribution scenario of Ref.~\cite{Clesse:2016vqa,Clesse:2017bsw} passes all the constraints in the stellar-mass range of interest here.

\

\textit{Conclusion:}  Three recent gravitational-wave observations, GW190425, GW190521 and GW190814, have attracted attention because they would involve compact objects in the so called lower and upper mass gaps and, moreover, none of them seem to have any significant spin.  We have shown that these properties, as well as the merging rates for these three events, are naturally explained if these objects are primordial black holes with a mass distribution imprinted by the thermal history of the Universe, at the time of the QCD epoch.  Two binary formation channels have been investigated, by tidal capture in PBH clusters or in the early Universe. The former seems to explain well the GW observations but the latter cannot explain at the same time the rates of GW190425, GW190521 and GW190814, as well as the rates inferred for almost equal-mass binaries around $30~\Msun$ detected by LIGO/Virgo.

The relatively simple analysis performed in this work provides new motivations for a detailed investigation by the LIGO/Virgo collaboration, using more advanced statistical techniques like Bayesian model comparison between PBH and stellar BH models, applied to spins~\cite{Fernandez:2019kyb,Garcia-Bellido:2020pwq}, masses and rates and based on the full upcoming catalog of events in the O1, O2 and O3 observing runs.  
If a primordial origin were to be definitely confirmed, these observations may revolutionize our understanding of the nature of DM, the origin of matter~\cite{Garcia-Bellido:2019vlf,Carr:2019hud} and the physics at play in the Early Universe.

\

\textit{Acknowledgements}. JGB acknowledges support from the Reserch Project PGC2018-094773-B-C32 [MINECO-FEDER], and the Centro de Excelencia Severo Ochoa Program SEV-2016-0597. The work of SC was supported by the Belgian Fund for Research F.R.S.-FNRS.

\appendix

\section{Rate suppression of early binaries}

\begin{figure*}[t]
    \centering
    \includegraphics[width=16cm]{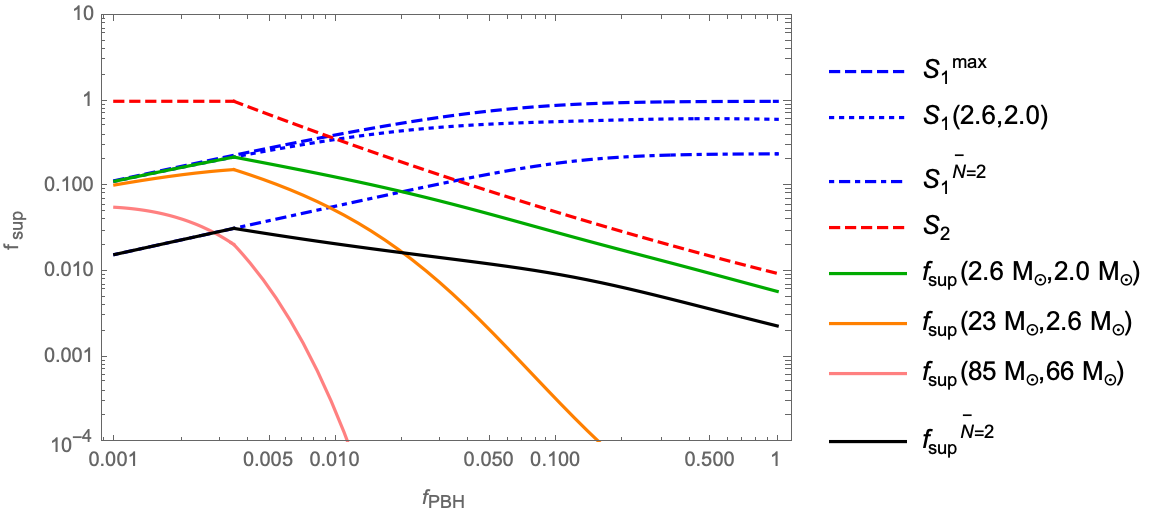}
    \caption{Merger rate suppression factor $f_{\rm sup}$ as a function of $f_{\rm PBH}$ for early binaries and the two contribution $S_1$ and $S_2$, in different cases.  Maximum value of $S_1$ from binary disruption by matter inhomogeneities and nearby PBHs ($S_1^{\rm max}$, dashed blue line).  Value of $S_1$ only taking into account disruption by PBHs from the peak with $m_1 = 2.6 M_\odot$ and $m_2 = 2.0 M_\odot$ (dotted blue line), and assuming $\bar N =2$ as in a monochromatic model (dotted dashed blue line). Value of the $S_2$ contribution from PBH clusters before matter radiation equality (dashed red line).  Total suppression factor assuming binary disruption by nearby PBHs from the QCD peak, with $m_1 = 2.6 M_\odot$ and $m_2 = 2.0 M_\odot$ (green line), $m_1 = 23 M_\odot$ and $m_2 = 2.6 M_\odot$ (orange line) and $m_1 = 85 M_\odot$ and $m_2 = 66 M_\odot$ (pink line) similar to the component masses of GW190425, GW190814 and GW190521.  In black, the most realistic case of a total suppression factor assuming $\bar N =2$ and leading to $f_{\rm sup} \simeq 2.5 \times 10^{-4}$ when $f_{\rm PBH} =1$, as in Fig.~\ref{fig:ratesprim}.
    }
    \label{fig:fsup}
\end{figure*}

Analytical prescriptions have been proposed in~\cite{Raidal:2018bbj,Hutsi:2020sol} to calculate the suppression factor $f_{\rm sup}(f_{\rm PBH}, m_1,m_2)$ that can be written as the product of two factors $S_1(f_{\rm PBH},m_1,m_2)$ and $S_2(f_{\rm PBH})$. They correspond to the rate suppression due to nearby PBHs or matter fluctuations and due to PBH clusters seeded by Poisson fluctuations, respectively.  These prescriptions have been compared with N-body simulations, but only in the cases of a monochromatic and a log-normal PBH mass distribution.   As we discussed below, one must be cautious when applying these prescriptions to our broader mass function, even if it exhibits a high and sharp peak at the solar mass scale from the QCD transition.

The first suppression factor is given by
\be
S_1 \approx 1.42 \left[ \frac{(\langle m_{\rm PBH}^2 \rangle/\langle m_{\rm PBH} \rangle^2)}{\bar N + C} + \frac{\sigma_{\rm M}^2}{f_{\rm PBH}^2}\right]^{-21/74} {\rm e}^{-\bar N}  \label{eq:S1}
\ee
that takes into account the binary disruption by either matter fluctuations with a (rescaled) variance $\sigma_{\rm M}^2 \simeq 0.005$ or by the number of nearby black holes $\bar N$ within a sphere around the binary whose radius is determined by the maximal comoving distance for a nearby PBH to fall onto the binary before matter-radiation equality.  It is estimated by
\be
\bar N = \frac{m_1+m_2}{\langle m_{\rm PBH} \rangle} \frac{f_{\rm PBH}}{f_{\rm PBH}+\sigma_M}~.  \label{eq:Nbar}
\ee
In Eqs.~(\ref{eq:S1}) and~(\ref{eq:Nbar}) the mean PBH mass and the corresponding variance are related to the mass function through
\bea
\langle m_{\rm PBH} \rangle & = &  \frac{\int m_{\rm PBH} {\rm d} n_{\rm PBH}}{n_{\rm PBH}} \nonumber \\  &= & \left[\int \frac{f(m_{\rm PBH})}{m_{\rm PBH}} {\rm d} \ln m_{\rm PBH} \right]^{-1} \\
\langle m_{\rm PBH}^2 \rangle & =& \frac{\int m_{\rm PBH}^2 {\rm d} n_{\rm PBH}}{n_{\rm PBH}} \nonumber \\
& = & \frac{\int m_{\rm PBH} f(m_{\rm PBH}) {\rm d} m_{\rm PBH}}{\int \frac{f(m_{\rm PBH}) }{m_{\rm PBH}}{\rm d} \ln m_{\rm PBH}}
\eea
where $n_{\rm PBH}$ denotes the total PBH number density.   The function $C$ encodes the transition between small and large $\bar N$ limits.  A good approximation is given by~\cite{Hutsi:2020sol}
\bea
C &\simeq& \frac{f_{\rm PBH}^2 \langle m_{\rm PBH}^2 \rangle} {\sigma_{\rm M}^2 \langle m_{\rm PBH} \rangle^2}  \nonumber \\
& \times & \left\{ \left[ \frac{\Gamma(29/37)}{\sqrt\pi} U \left( \frac{21}{74},\frac{1}{2}, 
\frac{5 f_{\rm PBH}^2}{6 \sigma_{\rm M}^2}\right) \right]^{-74/21}  -1 \right\}^{-1}
\eea
where $\Gamma$ is the Euler function and $U$ is the confluent hypergeometric function.

The second factor $S_2(f_{\rm PBH})$ comes from the binary disruption in early-forming clusters and can be approximated today by
\be
S_2 \approx \min \left(1,9.6 \times 10^{-3} f_{\rm PBH}^{-0.65} {\rm e}^{0.03 \ln^2 f_{\rm PBH}} \right).
\ee

We have computed $S_1$, $S_2$ and the resulting suppression factor $f_{\rm sup}$ for our mass function and for the mean masses of the three events GW190425, GW190814 and GW190521, as well as the maximal $S_1^{\rm max}$ obtained in the limit $\bar N \ll \min(C,1) $ that is independent of the two binary component masses.  They are represented on Fig.~\ref{fig:fsup}, as a function of $f_{\rm PBH}$.

One important difference with respect to the monochromatic or lognormal mass function is that the large number density of tiny black holes implies that $\langle m_{\rm PBH} \rangle \ll M_{\odot}$ and $\langle m_{\rm PBH}^2 \rangle / \langle m_{\rm PBH} \rangle^2 \ll 1$, even if $f(m)$ in this range is of order $10^{-2}$.  This implies that $\bar N \gg 1$, which leads to a huge exponential suppression of the merging rates.  These analytical prescriptions, when strictly applied to a broad mass function with thermal effects, thus leads to merging rates for early binaries that are much lower than the ones inferred from GW observations, much below the merging rates from PBH clusters.  However, the rate suppression is likely overestimated because PBHs that are much lighter than the binary components are likely not able to disrupt it.  Instead one could integrate the mass function over the QCD peak only.  By doing so, the suppression factor associated to GW190425 would be slighly below $S_2^{\rm max}$ and can be compatible with our benchmark choice, $f_{\rm sup} = 0.002 $.   Nevertheless, for PBH mergers with larger masses, the suppression is still quite efficient because $\bar N \gtrsim 1$.   It is therefore not possible to explain the rates of GW190814 and GW190521.  But again, it is difficult to know if PBHs from the QCD peak are able to disrupt more massive binaries.   Finally, one can consider only the disruption by nearby PBHs whose mass is similar to the mean of the binary component masses.  By doing so, one gets $\langle m_{\rm PBH} \rangle \sim (m_1+m_2)/2$ and $\langle m_{\rm PBH}^2 \rangle / \langle m_{\rm PBH} \rangle^2 \sim 1$.  In such a case, one gets $\bar N \approx 2$ (as in the monochromatic case) and the suppression factor obtained when $f_{\rm PBH} \simeq 1$ becomes independent of the mass, slightly below $S_2^{\rm max}$ depending on the exact value of $\bar N$, with $f_{\rm sup}$ between $10^{-3}$ and $10^{-2}$.  In particular, for $\bar N = 2$ and $f_{\rm PBH}= 1$ one gets $f_{\rm sup} \simeq 0.002$ that corresponds to our benchmark value.  This motivates our choice on a theoretical point of view.  However, one should keep in mind that there are still large uncertainties, related to the disruption by nearby PBHs.   

Even if the rates given by Eq.~\ref{eq:tauEB} are consistent with N-body simulations, there are still a series of uncertainties that may limit this analysis.   First, no N-body simulations have been performed in the case of our broad mass function with thermal features and so it is still possible that tiny or heavy PBHs far from the QCD peak additionally suppress these merging rates.  Second, Eq.~\ref{eq:tauEB} does not take into account the merging rates of the perturbed binaries that may become dominant when $f_{\rm PBH} \gtrsim 0.1$
~\cite{Vaskonen:2019jpv}, but there are not yet clear analytical prescriptions for binaires with non-equal masses.  Third, slightly different results and another possible dependence in $f_{\rm PBH}$ have been obtained in~\cite{Kocsis:2017yty} using analytical methods.   Fourth, it has recently been claimed in~\cite{Boehm:2020jwd} that subtle general relativistic effects may highly suppress this PBH binary formation channel, but this result has been disputed in~\cite{DeLuca:2020bjf,Hutsi:2021nvs}.   Given these limitations, one should remind that in general, early binaries can be impacted by their environment during the whole cosmic history, which changes significantly over time.  Strong claims relying on these merging rates are therefore probably still premature. Nevertheless, Eq.~(\ref{eq:tauEB}) probably gives a good estimate, at least in some regimes.

\section{Rate boost of late binaries}

\begin{figure}[t]
    \centering
    \includegraphics[width=8.7cm]{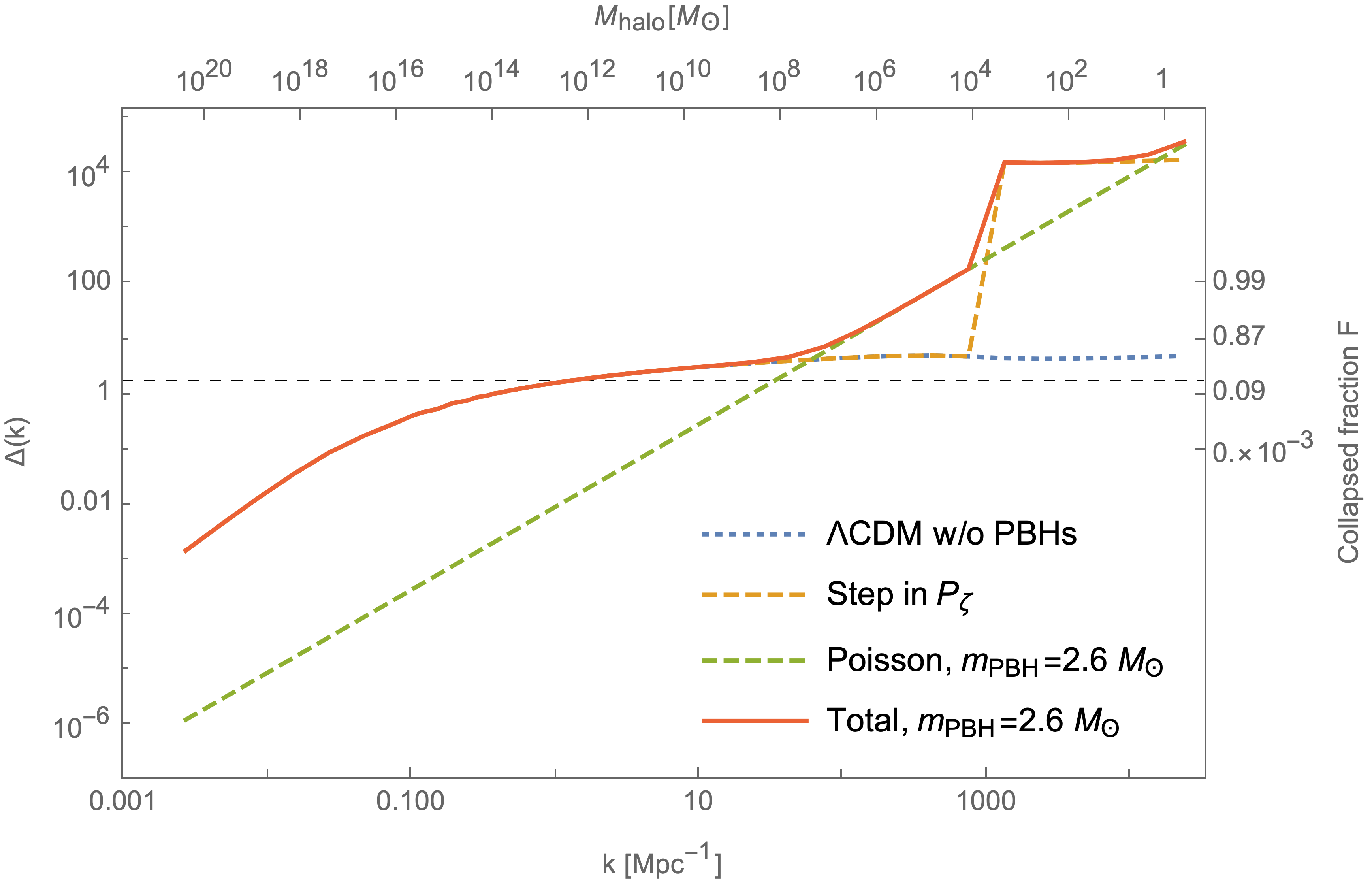}
    \caption{Adimensional (linear) density contrast $\Delta (k) $ today in the standard $\Lambda$CDM model with no PBHs (dotted blue line) and including PBHs with $f_{\rm PBH} =1$ at $m_{\rm PBH} =2.6 M_\odot$ (solid red line), including the unavoidable Poisson term in the matter power spectrum from the discrete nature of PBHs (dashed green line) and the effect of the power spectrum enhancement asssuming that the transition scale is $k_{\rm trans} = 10^3 {\rm Mpc}^{-1}$.  The horizontal dashed lines represents the critical threshold for halo formation $\delta_{\rm th}^{\rm halo} = 1.686$.   The upper x axis gives an estimation of the corresponding halo mass.  The right y axis shows the estimated halo collapsed fraction $F(M_{\rm halo})$.  Due to the Poisson term, one gets a natural clustering scale around halo masses of $10^6 - 10^7 M_{\odot}$, corresponding to ultra-faint dwarf spheroidals. }
    \label{fig:Delta}
\end{figure}

\begin{figure}[t]
    \centering
    \includegraphics[width=8.7cm]{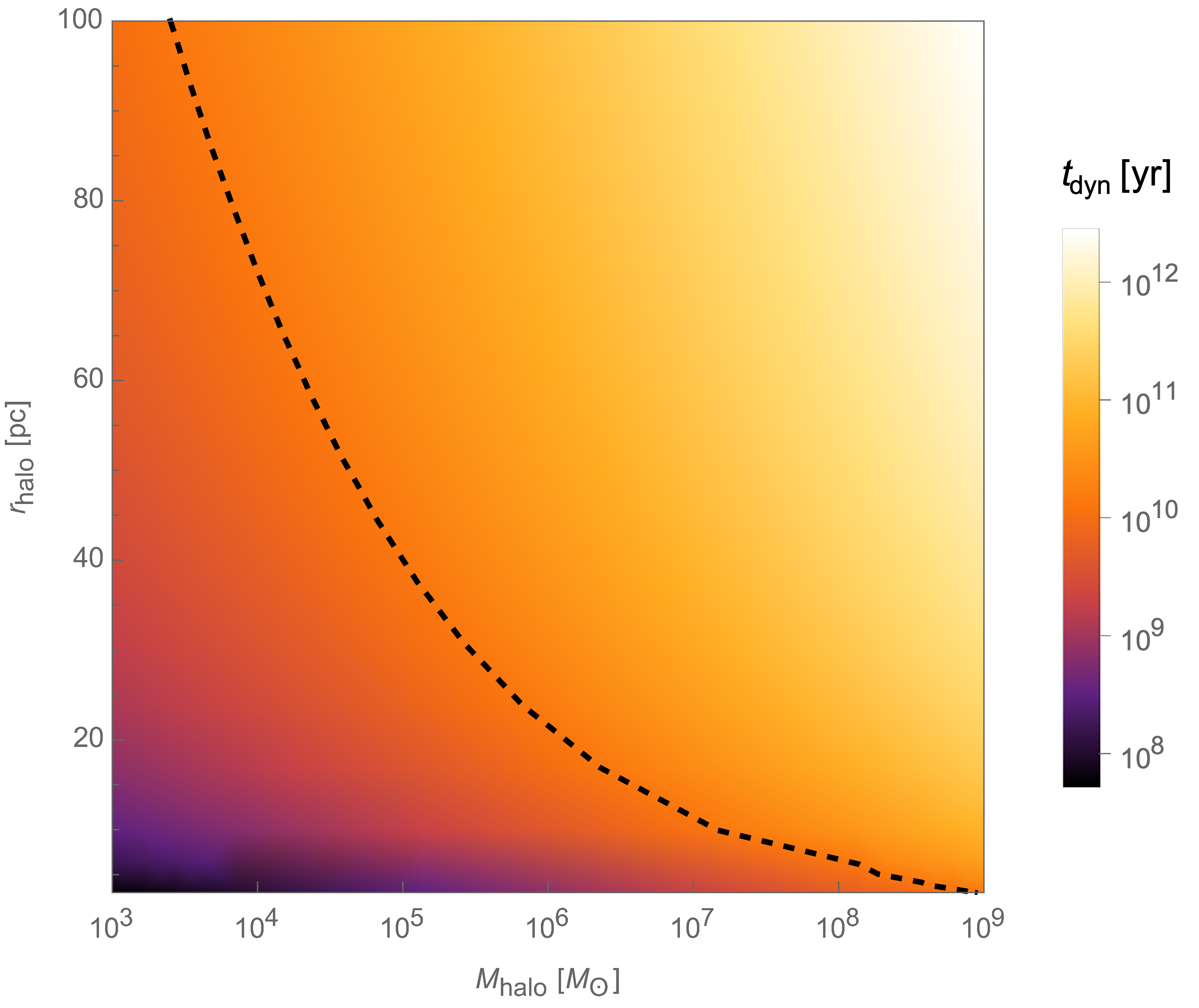}
    \caption{Dynamical heating time $t_{\rm dyn}$ needed for a subhalo of mass $M_{\rm halo}$ to reach a radius $r_{\rm halo}$ assuming a negligible initial size, $f_{\rm PBH} =1$ and $m_{\rm PBH} = 2.6 M_\odot$.  The dashed line corresponds to the age of the Universe.  Sub-clusters of mass below $10^6 M_\odot$ and of radius smaller than $\mathcal O(10)$pc quickly expand until they are completely diluted in their host cluster.  }
    \label{fig:tdyn}
\end{figure}

Regarding the merging rate of late binaries formed by tidal capture in clusters, one can wonder if the assumed value of $R_{\rm clust} \approx 420$ is consistent with the theoretical expectations for PBH clustering.  Indeed, if one considers the halo mass function as expected from the (extended) Press-Schechter formalism applied to the linear matter power spetrum of the standard cosmological model, one gets $R_{\rm clust} \approx \mathcal O(1-10)$~\cite{Bird:2016dcv} and merging rates that are lower than inferred from GW observations, even if $f_{\rm PBH} = 1$ with our extended mass function.

However, as pointed out in~\cite{Kashlinsky:2016sdv} in a different context, the discrete nature of PBHs induce an additional term in the matter power spectrum coming from Poisson fluctuations in the spatial distribution of PBHs at formation.  On small scales, this term dominates the matter power spectrum and unavoidably leads to the gravitational collapse of almost all small-scale perturbations into halos of masses up to $10^6 - 10^7 M_\odot$.  On Fig.~\ref{fig:Delta}, we have represented the  (linear) matter density contrast today
\be
\Delta (k) = \left( \frac{k^3 P(k)}{2 \pi^2}  \right)^{1/2}~,
\ee
where $P(k)$ is the total matter power spectrum, summing the linear power spectrum computed with the Boltzmann code \texttt{CLASS}~\cite{2011arXiv1104.2932L} with our primordial power spectrum and a transition between cosmological and PBH scales at $k = 10^3 {\rm Mpc}^{-1}$, and a Poisson constant term given (today) by~\cite{Kashlinsky:2016sdv}
\be
P_{\rm Poisson} \simeq 2 \times 10^{-2} \left( \frac{m_{\rm PBH}}{30 M_\odot}  \right) {\rm Mpc}^{3}~. 
\ee
For simplicity we assumed that all PBHs have the same mass $m_{\rm PBH} = 2.6 M_\odot$.  Due to the high QCD peak, summing over the whole mass function does impact the Poisson term only marginally.  To each scale one can associate a halo mass that roughly corresponds to the mass inside a fluctuation wavelength $\lambda = k / 2\pi$~\cite{Kashlinsky:2016sdv}
\be
M_{\rm halo} \simeq 1.15 \times 10^{12} \left( \frac{\lambda}{\rm Mpc} \right)^3 M_\odot~.
\ee
In the (extended) Press-Schechter formalism, the fraction of collapsed fluctuations into halos with a mass $M_{\rm halo} $ is given by
\be
F(M_{\rm halo}) = \rm{erfc} \left[ \frac{\delta_{\rm th}^{\rm halo}}{\sqrt 2 \sigma(M_{\rm halo})} \right]~,
\ee
where $\delta_{\rm th}^{\rm halo}\simeq 1.686$ is the overdensity threshold leading to the gravitational collapse and 
\be
\sigma^2(M_{\rm halo}) = \int \Delta^2(k) W(k) {\rm d \ln k}
\ee
in which we assume a Dirac-delta window function $W[k(M_{\rm halo})]$.  As shown in Fig.~\ref{fig:Delta}, one gets that $F(M_{\rm halo}< 10^6 M_\odot) $ is close to unity.

This fixes the natural clustering scale of PBHs around $10^6-10^7 M_\odot$.  Indeed, one can show that subhalos of smaller mass are dynamically unstable and expands until they are diluted in their host halo, with a typical dynamical heating time $t_{\rm dyn}$ obtained by solving~\cite{Brandt:2016aco}
\be
\frac{{\dd} r_{\rm halo}}{\dd t} = \frac{4 \sqrt 2 \pi G f_{\rm PBH} m_{\rm PBH} \ln (M_{\rm halo}/2 m_{\rm PBH} ) } {2 \beta v_{\rm vir} r_{\rm halo}}
\ee
where $r_{\rm halo}$ is the cluster radius, $v_{\rm vir}$ its virial velocity and $\beta \approx 10$ is a parameter depending on the halo profile.  We have represented $t_{\rm dyn}$ as a function of the subhalo mass and radius on Fig.~\ref{fig:tdyn}.   The perturbation length scale associated to halo masses below $10^6-10^7 M_\odot$ is smaller than the dynamically stable radius around matter-radiation equality.  Above this mass scale, halos have a larger radius and are dynamically stable at formation.    

This mass range for PBH clusters is particularly interesting for the interpretation of density perturbations on stellar tidal streams as arising from stochastic encounters with clumps of DM in the halo of our galaxy~\cite{Bovy:2016irg}. In the case of PBH clusters with mass between $10^5 - 10^7 \Msun$, the statistical methods developed in~\cite{Montanari:2020gcr} would clearly indicate their nature as the building blocks of DM halos in galaxies.

The last step is to estimate $R_{\rm clust}$.  For this purpose, we consider the merging rates with the explicit dependence in the PBH velocity and the averaged, enhanced local density contrast $ \delta^{\rm local} $ compared to the cosmological DM density.  From the rates of~\cite{Clesse:2016ajp}, one can identify
\be 
R_{\rm clust} = \frac{2 \pi \delta^{\rm local}  \Omega_{\rm M}^2  \rho_{\rm c} G }{c}    \left(\frac{85 \pi}{6 \sqrt 2 } \right)^{\frac{2}{7}} \left( \frac{c}{\sqrt{2} v_{\rm vir}} \right)^{\frac{11}{7}} \left(\frac{\rm yr}{\rm Gpc^3}\right)\,,
\ee
with $\delta^{\rm local} = 3 M_{\rm halo}/(4 \pi r_{\rm halo}^3  \rho_{\rm DM}^0)$ and a Virial velocity $v_{\rm vir} = \sqrt{G M_{\rm halo} /(2 r_{\rm halo})}$.  For clusters with $M_{\rm halo} = 10^6 M_\odot$ and $r_{\rm Halo} \approx 20$ pc, this gives $R_{\rm clust} \approx 100 $ while for $M_{\rm halo} = 10^7 M_\odot $ and  $r_{\rm Halo} \approx 10$ pc, one gets $R_{\rm clust} \approx 750$.   These values correspond both to the radius obtained through dynamical heating and to the observed critical size of ultra-faint dwarf galaxies.

Despite our crude assumptions, this theoretical estimate of $R_{\rm clust}$ is remarkably consistent with the value of $R_{\rm clust} \approx 420$ needed to explain the merger rates of GW190425, GW190814 and GW190521.  A refined analysis will nevertheless be useful in order to estimate more accurately and quantitatively the effective value of $R_{\rm clust}$, ideally using N-body simulations and including non-trivial effects~\cite{Trashorras:2020mwn}, such as PBH mass segregation in clusters, halo mass and velocity profiles, central intermediate mass black holes, cluster disruption by the host galaxy, etc.  

\bibliographystyle{apsrev4}

\bibliography{mainNotes.bib}

\end{document}